\documentclass[reprint,
 amsmath,amssymb,
 aps,
]{revtex4-2}

\usepackage{graphicx}
\usepackage{dcolumn}
\usepackage{bm}
\usepackage{xcolor}
\usepackage{mathptmx}
\usepackage{graphicx}
\usepackage{dcolumn}
 
\usepackage{bm}
\usepackage{amssymb}
\usepackage{comment}
\usepackage{hyperref}
\usepackage{physics}

\begin{document}

\title{Passive Cross-Basis Mode Transitions Along a Single Freely Propagating Bessel Beam}

\author{Henry P. Evans}
\author{Layton A. Hall$^{*}$}

\affiliation{Materials and Physics Applications-Quantum Division, Los Alamos National Laboratory, Los Alamos, NM 87545, USA}
\affiliation{$^*$Corresponding author: laytonh@lanl.gov}

\begin{abstract}
The transverse modal identity of a freely propagating optical beam is ordinarily fixed at the point of generation. We show that the conical angular spectrum of a Bessel beam establishes a one-to-one mapping between radial beam position and axial reconstruction distance. This mapping converts the radial aperture of a single static, phase-only spatial light modulator into a programmable longitudinal-mode register. By partitioning the modulator into independent annular regions, we encode discrete transverse modes at preselected axial positions. We demonstrate this principle with programmable ring-lattice fields of axially varying site number, and with passive transitions that sequence through Bessel, Bessel vortex beam, Hermite-Gaussian-Bessel, and Airy caustic modes within a single beam, without dynamic modulation or cascaded optical elements.
\end{abstract}


\maketitle

Transverse spatial modes of optical beams, including Hermite--Gaussian (HG), Laguerre--Gaussian (LG), Airy, and Bessel families, constitute a rich set of orthogonal state spaces for encoding classical and quantum information, driving light--matter interactions, and structuring optical potentials~\cite{Allen1992, Forbes2021, Rubinsztein2017, Andrews2013}. Each family is characterized by a distinct underlying spatial symmetry: Cartesian for HG, cylindrical for LG, caustic for Airy, and propagation-invariant for Bessel~\cite{Durnin87PRL, Siviloglou2007}. These modal families are usually treated as distinct bases and are not normally interconverted by simple propagation of a single prepared mode.

Transitions between modal families are possible with dedicated optical elements. Astigmatic mode converters continuously deform the Gaussian envelope to map HG onto LG modes~\cite{Beijersbergen1993, Allen1992}. Multi-plane light conversion and cascaded diffractive or metasurface elements implement more general unitary transformations between transverse modes~\cite{Devlin2017}, and dynamically reconfigurable SLMs can synthesize arbitrary wavefronts at selected planes~\cite{Bolduc2013, Maurer2011}. In each case, the modal identity at a given plane is set by an active element or by a cascade, not by a passive feature of free-space propagation itself.

A complementary line of work tailors the longitudinal degree of freedom while holding the transverse symmetry class fixed. Frozen waves sculpt axial intensity by superposing co-axial Bessel beams with prescribed wavenumber spectra~\cite{ZamboniRached2005,Mendonca2025}; twisted frozen waves extend this to coherent axial transfer of OAM structure~\cite{TwistedFW2026}; and the rotatum of light reveals additional axial structure in the OAM current~\cite{Rotatum2025}. A recent perspective consolidates this rapidly growing area but explicitly scopes it to continuous parameters, including intensity, OAM, polarization, and wavelength within a fixed transverse-mode family~\cite{Willner2025LongitudinalPerspective}. Bessel–Gaussian superpositions with discrete longitudinal wavenumbers produce beams whose topological charge $\ell$ steps during propagation~\cite{dorrah2018}, metasurfaces realize longitudinal variation of polarization and OAM~\cite{dorrah2022}, and different-cone-angle Bessel–Gauss superpositions generate optical-needle arrays with independent axial intensity and vortical structure~\cite{Orlov2023BesselArray}.

Two limitations are common: modal evolution stays within one transverse symmetry class (frozen waves, twisted frozen waves, $\ell$-stepping, needle arrays) or, when crossing classes, needs polarization-sensitive subwavelength elements, cascaded optics, or dynamic modulation (mode converters, MPLC, metasurfaces, reconfigurable SLMs). Neither enables passive, deterministic transitions between incompatible transverse symmetry classes—Cartesian, cylindrical, caustic, and propagation-invariant—along a single beam.

Here, we demonstrate that the conical angle of Bessel beams provides a natural physical mechanism for cross-modal changes with propagation. The one-to-one geometrical correspondence between radial beam position and axial reconstruction distance, which is inherent to axicon-generated Bessel fields~\cite{Durnin1987, McGloin2005}, transforms the radial coordinate of a phase-only SLM into a programmable longitudinal modal register.
The novelty of this work is therefore not the finite-aperture axicon mapping itself, which is well established~\cite{Durnin87PRL}, but its use as a passive longitudinal modal register. That is, independent annular phase holograms on a single static, phase-only, polarization-insensitive SLM encode distinct transverse-mode families at preselected axial planes, and the symmetry class of the beam changes discretely with $z$.
Partitioning the SLM aperture into independent annular regions encodes different transverse modes, creating passive free-space beams with evolving modal and topological states during propagation without dynamic modulation or cascaded elements. We demonstrate programmable ring-lattice fields with axial site variation and discrete cross-basis transitions, sequentially switching through Bessel, vortex, Hermite-Gaussian-Bessel, and Airy modes within one beam. These results establish the longitudinal propagation coordinate as an addressable modal degree of freedom with direct implications for volumetric optical manipulation, axially multiplexed communications~\cite{Wang2012, Bozinovic2013}, and three-dimensional photonic state preparation~\cite{Forbes2019}.

\begin{figure}[t!]
    \centering
    \includegraphics[width=7.6cm]{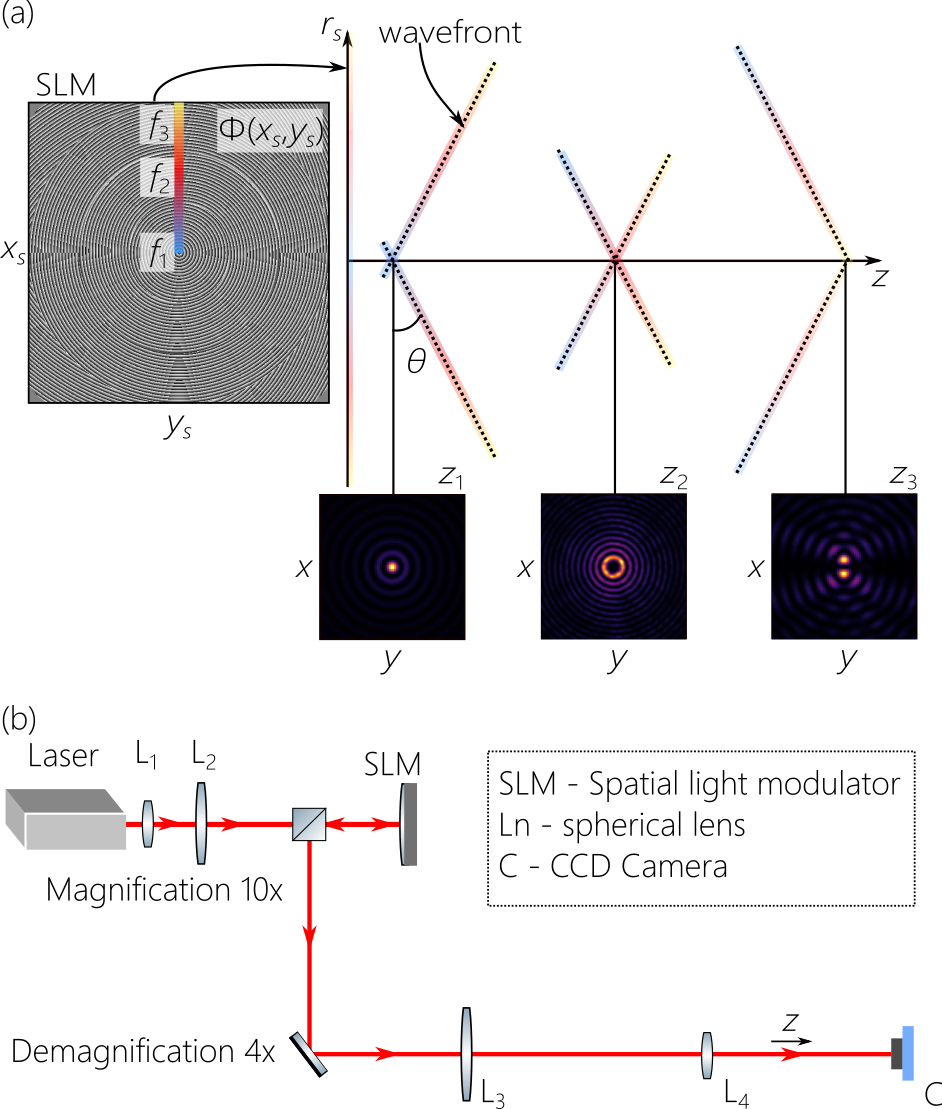}
    \caption{Principle and setup. (a) A phase-only axicon $\Phi(x_s,y_s)$ is partitioned into concentric annuli $f_m$; the Bessel cone angle $\theta$ maps each radius $r_{m,s}$ to a unique reconstruction plane $z_m = k\,r_{m,s}/k_r$, so distinct transverse modes appear at $z_1,z_2,z_3$. (b) A 635~nm laser is expanded $4\times$ (L$_1$: $f_1 = 50$~mm,L$_2$: $f_2 =200$~mm), reflected from the SLM, and relayed through a $4f$ system (L$_3$: $f_3 = 400$~mm, L$_4$: $f_4=100$~mm; $4\times$ demagnification) onto a translating CCD (C).}
    \label{fig:Theory}
\end{figure}

\textbf{Principle of Longitudinal Modal Encoding} --- An ideal Bessel beam is formed by the interference of plane waves distributed on a cone of fixed half-angle $\theta$ [Fig. \ref{fig:Theory}(a)]. In a finite-aperture axicon geometry, the transverse wave vector $k_r = k\sin\theta$ fixes the cone angle and $k$ is the wavenumber, while the launch radius $r_s$ determines the axial position at which that annular contribution reconstructs, $z \simeq r_s k/k_r$. This geometrical mapping between annular launch radius and axial reconstruction position is the physical foundation of the method presented here.

For a conventional zero-order Bessel beam generated by a phase-only axicon, the SLM phase $\Phi$ is
\begin{equation}
    \Phi(r_s, \varphi) = k_r r_s,
    \label{eq:axicon}
\end{equation}
where $r_s$ is the radial coordinate on the SLM and $k_r = k\sin\theta$ is the radial spatial frequency. To encode longitudinally programmable transverse structure, we partition the axicon phase into $M$ independent concentric annular regions, each assigned a distinct phase modulation $f_m(r_s, \varphi)$:
\begin{equation}
    \Phi(r_s,\varphi) = k_r r_s + \sum_{m} f_m(r_s,\varphi_s)
    \,\Pi\!\left(\frac{r_s - r_{m,s}}{\Delta r}\right),
    \label{eq:segmented}
\end{equation}
where $\Pi(\cdot)$ is the rectangle function, $r_{m,s}$ is the central radius of the $m$th annulus, and $\Delta r$ is its radial width. Each annular region acts as an independent phase hologram whose encoded transverse structure is reconstructed only at its designated axial position. This can be represented in Fig. \ref{fig:Theory}(a) on the SLM pattern where the varying colors represent the different regions of $f_m$.

\begin{figure}[t!]
    \centering
    \includegraphics[width=7.6 cm]{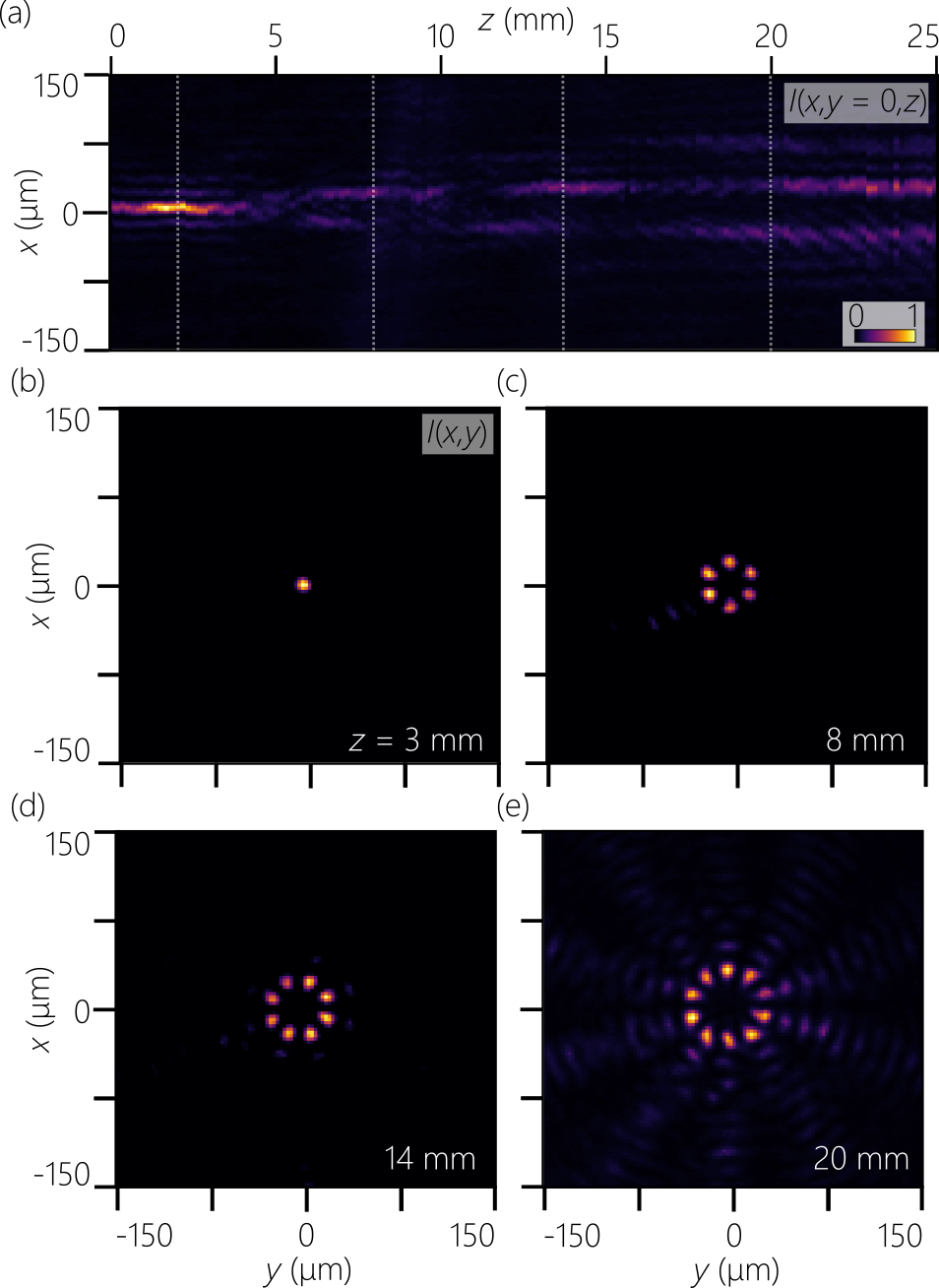}
    \caption{Axially programmable ring-lattice fields. (a) Measured longitudinal intensity cross-section $I(x, y=0, z)$  over 25~mm of free-space propagation; dashed lines mark the four axial observation planes. Transverse intensity profiles $I(x,y)$ at (b)~$z = 3$~mm ($N=1$),  (c)~$z = 8$~mm ($N=6$), (d)~$z = 14$~mm ($N=8$), and (e)~$z = 20$~mm ($N=12$),  demonstrating that the number of ring-lattice sites is independently programmable at each axial depth via the annular SLM geometry. All panels share the normalized intensity scale shown in~(a).}
    \label{fig:ring_lattice}
\end{figure}

The mapping between SLM radial coordinate and axial reconstruction distance follows from the finite-aperture axicon geometry. For a fixed transverse wave vector $k_r = k\sin\theta$, an annular zone centered at radius $r_{m,s}$ reconstructs near the axial plane $z_m$ determined, in the paraxial limit, by
\begin{equation}
    \frac{r_{m,s}}{z_m} \simeq \frac{k_r}{k},
    \label{eq:mapping}
\end{equation}
where $k = 2\pi/\lambda$ is the optical wavenumber and $z_m$ is the axial plane at which the $m$th encoded mode is reconstructed. Thus, the transverse wave vector fixes the cone angle, while the annular radius sets the axial address. The axial extent $\Delta z$ over which each encoded mode persists is set by the radial thickness $\Delta r$ of its annulus,
\begin{equation}
    \Delta r = \frac{k_r \, \Delta z}{k},
    \label{eq:extent}
\end{equation}
\begin{figure*}[t!]
    \centering
    \includegraphics[width=15 cm]{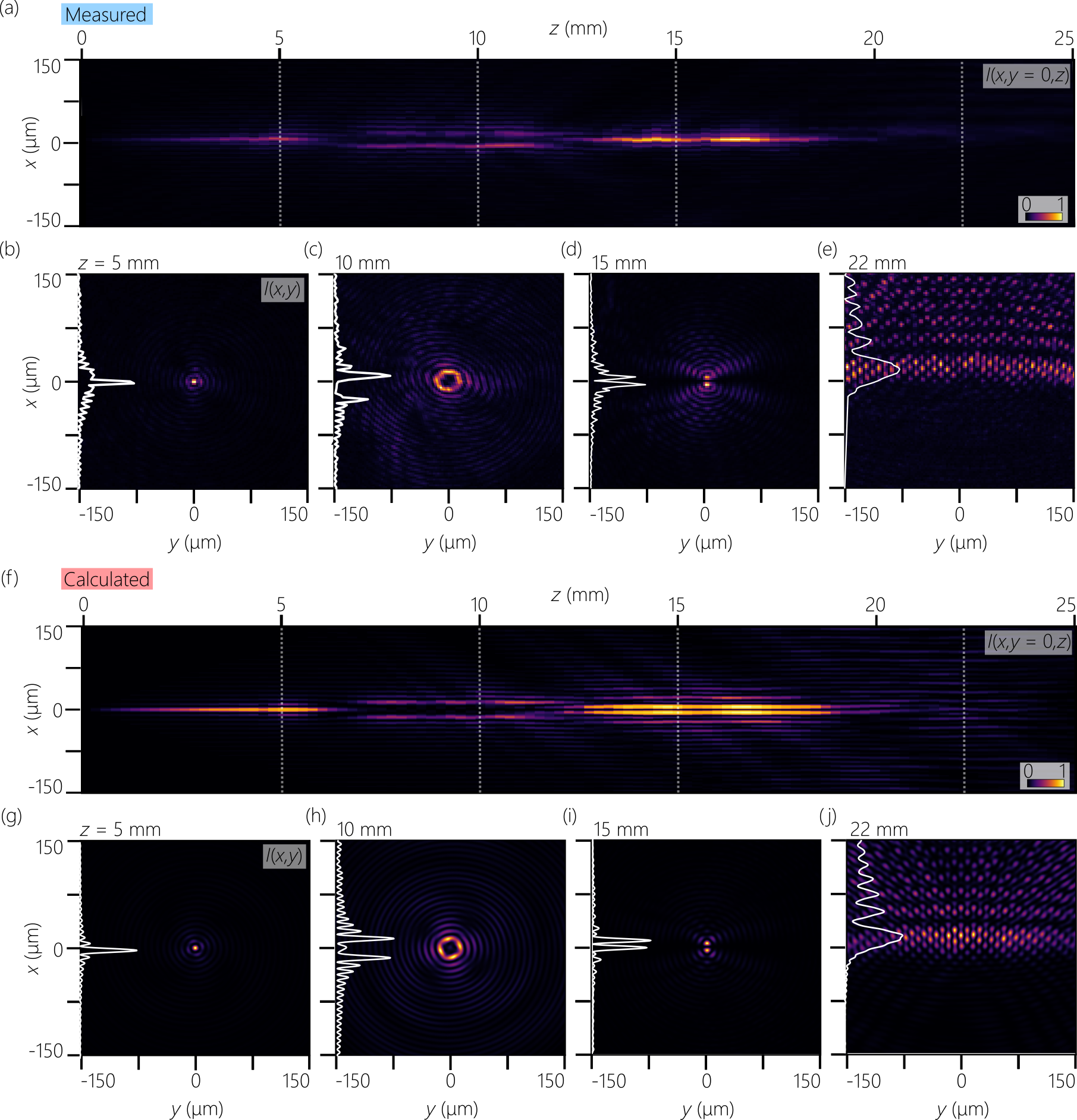}
    \caption{Passive cross-basis modal transitions along a single freely propagating beam encoded by a partitioned, phase-only SLM. (a) Measured longitudinal intensity cross-section $I(x, y=0, z)$ over 25~mm of free-space propagation; dashed lines mark the four axial observation planes. Transverse intensity profiles $I(x,y)$ measured at (b) $z = 5$~mm [zero-order Bessel-Gaussian], (c) $z = 10$~mm [Bessel vortex, $\ell = 4$], (d) $z = 15$~mm [HGB], and (e) $z = 22$~mm [Airy]. (f) Corresponding calculated longitudinal cross-section and (g--j) calculated transverse profiles at the same axial planes. White curves in (b--e) and (g--j) show intensity line profiles along the $x$-axis. }
    \label{fig:cross_basis}
\end{figure*}
so that both the axial position and longitudinal duration of each modal state are independently programmable through the annular geometry. This is physically represented in  Fig. \ref{fig:Theory}(a) where the wavefront is tilted uniformly, but the colors vary over the extent of the wavefront, representing the different regions of $f_m$. This mapping is purely geometrical in origin and places no restriction on the functional form of $f_m$; that is, any transverse mode structure expressible as a phase pattern, including HG, LG, Airy,  and Bessel modes, can be assigned to an independent axial register.  The propagation direction thereby functions as a longitudinal modal degree of freedom, encoded entirely within a single static,  phase-only SLM.

\textbf{Experimental Results} --- We demonstrate longitudinal modal encoding [Fig.~\ref{fig:Theory}(b)] using a continuous-wave laser at $\lambda = 635$~nm (Thorlabs HLS635) expanded to a 10~mm beam diameter and directed at normal incidence onto a phase-only SLM (Meadowlark HSP1K-488-800-PC8). The modulated beam is retroreflected, passed through a 50:50 beam splitter, and relayed through a $4f$ imaging system with focal lengths $f_3 = 400$~mm and $f_4 = 100$~mm, yielding a $4\times$ demagnification of the SLM plane onto the propagation axis. The output intensity is recorded transversely at successive axial positions using a CCD camera (Allied Vision 1800-U-500c) during axial scanning, from which the longitudinal cross-section $I(x, y=0, z)$ is reconstructed. In all scenarios, consider the spatial frequency of the base Bessel beam is $k_r = 0.2$~rad/$\mu$m.

As a first validation of the encoding principle, we partition the SLM aperture into concentric annular zones, each encoding a ring-lattice phase pattern with a distinct site number $N$. Fig.~\ref{fig:ring_lattice}(a) shows the measured longitudinal intensity cross-section over 25~mm of free-space propagation, confirming that each encoded mode is reconstructed at its designated axial position. Transverse profiles at $z = 3$~mm ($N=1$), $z = 8$~mm ($N=6$), $z = 14$~mm ($N=8$), and $z = 20$~mm ($N=12$) are shown in Figs.~\ref{fig:ring_lattice}(b--e), demonstrating that the number of lattice sites and therefore the azimuthal topology of the intensity landscape is independently programmable at each axial depth through the annular SLM geometry alone. We find that the number of points in each ring exactly matches the expected result. The full axial evolution is shown in Visualization~1.

Having confirmed that both axial position and azimuthal topology are independently programmable through the annular geometry, we turn to the central result: passive cross-basis modal transitions between fundamentally distinct transverse symmetry classes. For the cross-basis modal transition experiment, four annular zones are assigned phase patterns corresponding to a zero-order Bessel-Gaussian, a Bessel vortex of charge $\ell = 4$, an HGB, and an Airy caustic, respectively. Fig.~\ref{fig:cross_basis}(a) shows the measured longitudinal cross-section, where distinct modal features are visible at each of the four axial planes marked by dashed lines at $z = 5$, 10, 15, and 22~mm. The Airy zone is generated by spectrally convolving an 
Airy--Gaussian kernel with the base Bessel angular spectrum, imprinting caustic acceleration onto the radial Bessel distribution. The tight Gaussian confinement in $y$ renders the field effectively one-dimensional in its Airy character, closely approximating a canonical Airy--Gaussian beam.
The corresponding transverse profiles in Figs.~\ref{fig:cross_basis}(b--e) confirm the sequential reconstruction of each modal class at its designated depth. Figs.~\ref{fig:cross_basis}(f--j) show the calculated longitudinal cross-section and transverse profiles, which are in good agreement with the measurements. 
We find the intensity fidelity with the expected intensity profile to be 96\%, 84\%, 94\%, and 91\% in each plane, where the overlap is the normalized integral of $\left[\iint \sqrt{I_m(x,y)\,I(x,y)}\,dx\,dy\right]^{2}/\iint I_m(x,y)\,dx\,dy \;\cdot\; \iint I(x,y)\,dx\,dy$. In addition, crosstalk between regions is $<5\%$ in each region, where we define it as the fraction of energy that leaks into other regions' axial slabs when only that band's SLM zone is illuminated.
The passive generation of these cross-basis transitions, which span mutually incompatible Cartesian, cylindrical, caustic, and propagation-invariant symmetry classes from a single static, phase-only optical element, without dynamic modulation or cascaded components, constitutes the central experimental result of this work. The full axial evolution is shown in Visualization~2.

\textbf{Discussion} --- The results demonstrate that the longitudinal propagation coordinate of a freely propagating beam can function as an addressable modal degree of freedom, programmed entirely from a single static, phase-only SLM. The geometric one-to-one correspondence of Eq.~(\ref{eq:mapping}) imposes no restriction on the functional form of the encoded phase patterns, meaning that any transverse mode expressible as a phase hologram which includes higher-order modes, fractional-charge vortices, and arbitrary caustic beams can in principle be assigned to an independent axial register within the same framework.

The ability to vary modal identity and topological state along a single beam axis has direct consequences for volumetric optical manipulation. In optical tweezer arrays, different particle species or trapping geometries can be simultaneously addressed at distinct axial depths without mechanical refocusing or dynamic modulation~\cite{Rubinsztein2017}. The ring-lattice result shows that the number of lattice sites, and therefore the symmetry of the trapping potential, is independently programmable at each depth, enabling spatially multiplexed potentials with axially varying topology within a single static beam. For free-space communications, axial modal multiplexing offers a complementary degree of freedom to transverse OAM multiplexing~\cite{Wang2012, Bozinovic2013}, potentially supporting range-selective receivers without spatial demultiplexing optics. The passive generation of discrete cross-basis transitions within a single beam suggests a route to preparing three-dimensional photonic states without cascaded elements~\cite{Forbes2019}, with an encoding principle that is wavelength-scalable and compatible with ultrafast pulse fronts.

\textbf{Conclusion} --- We have shown that the conical angular spectrum of Bessel beams maps radial beam position to axial reconstruction distance, turning a single static phase-only SLM into a programmable longitudinal modal register. Using this principle, we realized passive cross-basis transitions among Bessel, Bessel vortex, Hermite--Gaussian--Bessel, and Airy caustic modes within a single freely propagating beam, with intensity fidelity of 84--96\% and crosstalk below 5\%.   \\

\noindent\textbf{Funding}
Los Alamos National Laboratory LDRD program grant 20251140PRD1. \\

\noindent\textbf{Acknowledgments}
We thank M. Martin for the equipment and space, and D. Borenstein for proofreading. \\ 

\noindent\textbf{Supplemental Materials}
See Visualization~1 and~2 for a supplementary movie showing the CCD-recorded intensity profile during axial scanning through the ring-lattice and the cross-basis modal transitions.\\

\noindent\textbf{Disclosures}
The authors declare no conflicts of interest.\\

\noindent\textbf{Data availability}
Data underlying the results presented in this paper is available upon reasonable request.


\bibliography{diffraction}

\end{document}